# Analysis of Controlled Photon Storage Time Using Phase Locking by Atomic Population Transfer

Byoung Seung Ham
*Center for Photon Information Processing, School of Electrical Engineering, Inha University, 253 Yonghyun-dong, Nam-gu, Incheon 402-751, S. Korea*
bham@inha.ac.kr

Atomic population transfer in an inhomogeneously broadened optical medium is analyzed for on-demand photon storage-time control in both atomic frequency comb (AFC) and phase locked echoes. In AFC the photon storage mechanism belongs to the conventional three-pulse photon echoes, while the storage time control is limited by conventional two-pulse photon echoes. ©2010 Optical Society of America

*OCIS codes:* 270.1670, 190.4380, 300.6240.

In an inhomogeneously broadened two-level optical system, resonant optical fields can induce collective atomic coherence, where each atom's coherence magnitude depends on the population change resulting from the inhomogeneous detuning. The excitation of atomic coherence by a single optical pulse brings all interacting atoms are into in-phase, but they quickly diphase at the end of the light pulse due to each predetermined phase velocity by the inhomogeneous detuning. When these atoms are flipped over across the resonance frequency, for example by a $\pi$ optical pulse as shown in conventional photon echoes [1,2] or by a Stark field in gradient echoes [3], all atoms' phase evolutions become reversed resulting in a collective coherence burst as a photon echo. To slow or even momentarily stop the atomic coherence decay, an independent control light can be used to transfer excited atoms into an auxiliary ground spin state, where the spin state is robust against the decay processes. This on-demand atom population transfer, or optical deshelving leads to coherence conversion from an optically excited state to a robust spin state.

    In an ultraslow light regime, atomic population transfer-based coherence conversion has been experimentally demonstrated for coherence swing between optical and spin states [4]. In the coherence swing, optical phases of all atoms before and after the control pulse must be in phase, so that nondegenerate four-wave mixing processes can determine such outcomes as matched pulse generation [5]. In the nondegenerate four-wave mixing or phase conjugate processes using the collective coherence, the atomic phase must be within the coherence limit [5,6]. This requirement of coherence between consecutive optical pulses can be removed if a rephasing process is involved, such as in photon echoes, where the rephasing is a time reversed process to restore the initial collective coherence leading to photon echoes [1,2,7]. Thus, the rephasing-based coherence conversion process relaxes the strict in-phase condition, so that the control pulses do not need to be coherent with each other or with the initial excitation pulse(s). Such an optical deshelving-based coherence conversion has been experimentally demonstrated in atomic frequency comb (AFC) echoes [8] and phase locked echoes [9]. In these deshelving processes, the rephasing of atoms stops due to lack of population decay, but with a gain of phase shift: A control pulse induces a $\pi/2$ phase shift for each $\pi$ pulse area [10].

    In this Letter we discuss atomic coherence transfer via optical deshelving between optical (excited) and spin (auxiliary ground) states using independent control lights, where the purpose of coherence population transfer is to minimize or to halt the coherence decay process of the excited atoms. Such a population transfer technique originated in incoherent spontaneous emission process [11] has been adapted by AFC to overcome shorter photon storage time [12]. Specifically here, AFC echoes are analyzed to explain the photon storage mechanism with a phase recovery condition. For this, the AFC echo is compared with the phase locked technique to illustrate similarities and differences.

    Figure 1(a) shows an energy level diagram of an optical medium for the analysis of modified photon echoes using quantum coherence conversion process. Light P induces atomic coherence between states |1> and |3>. Light B1 is a deshelving control pulse to transfer atoms from state |3> to state |2>, where quantum coherence $\rho_{13}$ excited by Ps is converted into spin coherence $\rho_{12}$ via population transfer only. Light B2 is a match pulse to reverse the process returning both atoms' phase and population on state |3>. Figure 1(b) shows the interacting light pulse

sequence, where the deshelving control pulses B1 and B2 function to extend storage time due to longer coherence decay of $\rho_{12}$.

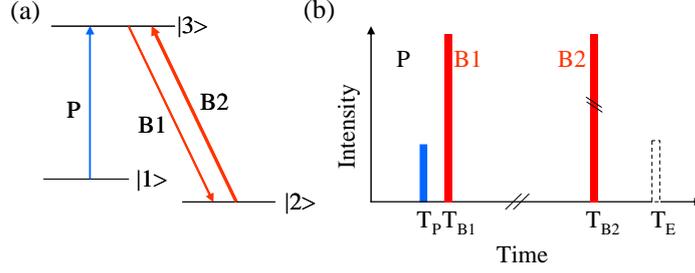

**Fig. 1.** (a) Energy level diagram and (b) pulse sequence.

For numerical simulations, time-dependent density matrix equations for Fig. 1 are derived from:
$$i\hbar\dot{\rho} = [H,\rho] + \text{decay terms}, \qquad (1)$$
where H is Hamiltonian and ρ is a density matrix operator: $\rho = |\Psi\rangle\langle\Psi|$; |Ψ> is a state vector. From Eq. (1) nine coupled equations are obtained and numerically solved without any assumption. For simplification, the spin decay rates $\Gamma_{12}$ and $\gamma_{12}$ are set to zero. For visual effects of population decay in AFC, a greater optical decay rate $\Gamma_{3j}$ (j=1,2) is assumed.

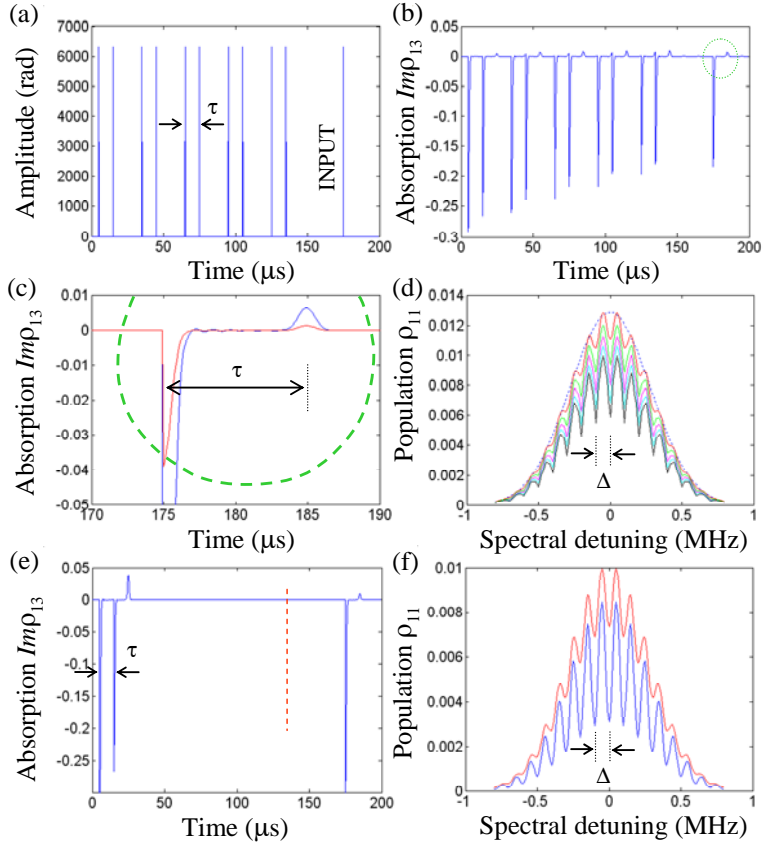

**Fig. 2.** (a) ACF echo pulse sequence. Each pulse turns on at 5, 15, 35, 45, 65, 75, 95, 105, 125, 135, and 175 μs. Each pulse duration is 100 ns. (b) and (c) Absorption vs. time for (a). The dotted circle indicates the AFC echo. (d) Population $\rho_{11}$ vs. atom detuning. t=16 (red), 46 (green), 76 (magenta), 106 (cyan), 136 μs (black). Dotted curve represents initial spectral distribution at t=0. (e) Conventional three-pulse photon echo. (f) Population $\rho_{11}$ vs. atom detuning. Blue: t=16 μs; Red: t=136 μs. For all, $\Gamma_{31}=\Gamma_{32}$=20 kHz (population decay rate); $\gamma_{31}=\gamma_{32}$=30 kH (phase decay rate); $\Gamma_{21}=\gamma_{21}$=0; atoms are Gaussian distributed: FWHM=680 kHz. Initial population is $\rho_{11}$=1;$\rho_{22}=\rho_{33}$=0.

In Fig. 2 the AFC echo is calculated and compared with a conventional three-pulse photon echo to analyze its storage mechanism. Figure 2(a) shows the AFC pulse sequence of P without use of control pulses. For AFC preparation, a light pulse train composed of five two-pulse pairs is applied long before the last single pulse, INPUT. The population decay time is ~10 μs ($\Gamma_{31}=\Gamma_{32}=20$ kHz), so that there is no atom left on state |3> when the INPUT pulse is applied. Figure 2(b) shows the result of Fig. 2(a) for absorption, $Im\rho_{13}$. Each light pulse area Φ is set to be small: Φ=π/5. The AFC echo amplitude (dotted circle) is proportional to the magnitude of INPUT power in a weak field limit as shown in Fig. 2(c). This is because information is stored in the spectral gratings made by the preparation pulse sets, and the INPUT pulse scatters off the grating to generate an echo (will be explained below). Figure 2(d) shows the ground state atom population $\rho_{11}$ resulting from each preparation pulse set as a function of atom detuning. The spectral spacing Δ is determined by the temporal spacing τ of the preparation pulse set: Δ=1/τ. Because all pulse sets have the same temporal separation τ, the spectral grating sharpens as the pulse sets increase. Eventually with many pulse sets, half of the atoms are left in state |1>, and half are dumped into state |2> via a spontaneous emission process. This population loss into state |2> indicates 50% coherence loss for each pulse set in normal photon echoes. In AFC, however, extremely weak light or single photons are used as INPUT (or read-out pulse), and thus this coherence loss is negligible [12]. Figure 2(e) as a conventional three-pulse photon echo is to compare with Fig. 2(b). For this only the first preparation pulse set is used with a π/2 pulse area to maximize echo efficiency [2]. No phase locking assumption is made in order to intentionally show the two pulse echo marked by "1" as compared with the three pulse echo marked by "2." The echo magnitude decrease is due to population decay rate $\Gamma_{3j}$ (j=1,2). Figure 2(f) shows the corresponding spectral grating of Fig. 2(e): The red curve is for t=136 μs [red dotted line of Fig. 2(e)], while the blue curve is for t=16 μs, immediately following the second pulse in Fig. 2(e). The grating sharpness decreases due to population decay, $\Gamma_{31}$. Thus, Fig. 2 proves that the AFC echo belongs to the three-pulse photon echoes, where information is stored in the spectral grating. Unlike the explanation in Ref. 8, the AFC echo is a retrieval of the spectral grating, where INPUT is not the origin of the echo but plays as a read-out pulse. If INPUT is the same as the preparation pulse set or single photons, however, the AFC echo can be treated as INPUT.

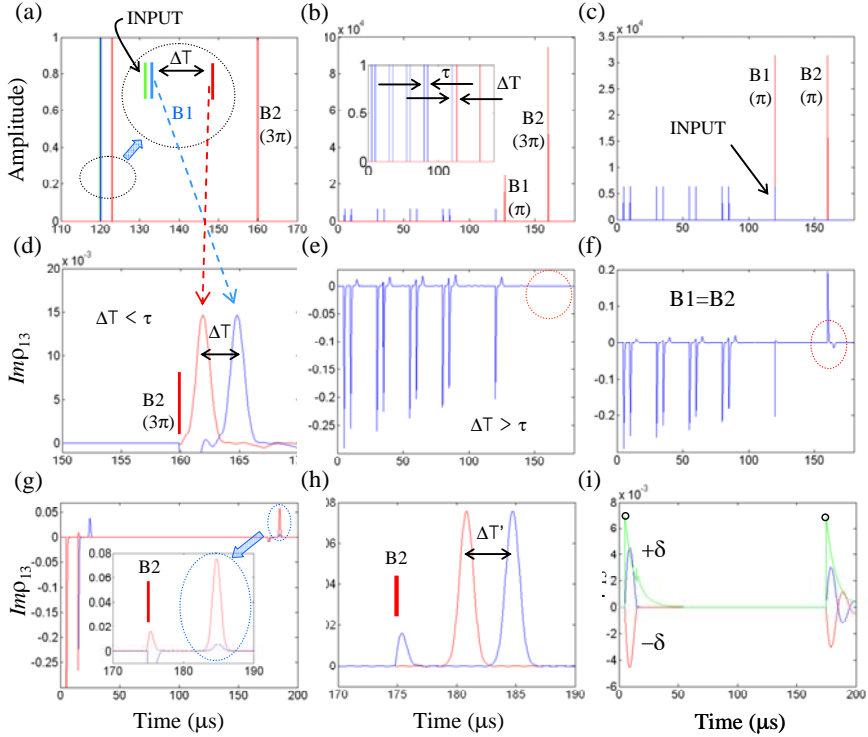

**Fig. 3.** (a) and (b) AFC echo pulse sequence with control pulses B1 and B2 for ΔT<t and ΔT>t, respecively. Pulse area of B1 and B2 satisfies π and 3π. (c) For (a) but with identical control pulses whose pulse area is π. (d)-(f) Numerical simulations of (a)-(c), respectively. (g) and (h) Phase locked echo corresponding to (a) and (d). (i) Individual atom phase evolution of (g) for δ=40 kHz; Green curve is for $\rho_{33}$.

Figure 3 represents storage time extension using deshelving control pulses B1 and B2. Temporal separation τ of each pulse set is 5 μs. Figures 3(a)-3(c) shows the pulse sequence for Figs. 3(d)-3(f), respectively for different conditions. In Figs. 3(a) and 3(d), the longer the delay of B1, the shorter the delay of the echo: The sum of the B1 delay and the echo delay is always the same as τ. If the B1 delay is longer than τ, however, the echo disappears as shown in Fig. 3(e). Figure 3(f) represents the phase locking condition of AFC in Ref. 8, where usage of identical deshelving control pulses violates the phase locking condition resulting in no echo [10]: This contradictory phenomenon has been explained with population leakage in a dilute medium [13]. Figures 3(g) and 3(h) show the phase locked echo corresponding to the AFC echo in Figs. 3(a) and 3(d). For these, Fig. 2(e) is modified: Right after the second pulse, B1 (π) is added and the third pulse is replaced by B2 (3π) according to the phase recovery condition [10]: $\Phi_{B1}+\Phi_{B2}=4\pi$. The blue one is the AFC echo in Fig. 3(d) for τ=10 μs: The weak AFC echo is due to population decay-caused coherence loss. The B1 delay-dependent echo position in Fig. 3(h) shows exactly the same physics as in Fig. 3(d) proving that the storage extension mechanism of the AFC echo is based on the same rephasing process as in the phase locked echoes. Figure 3(i) represents the rephasing mechanism of Fig. 3(g) in two-pulse photon echoes. With these phase evolutions, the phase of symmetrically δ detuned atoms is swapped (or rephased) by the second π pulse. Thus, the AFC echo uses spectral gratings for quantum optical storage, while the storage extension mechanism follows the rephasing process as in the phase locked echoes. The deshelving pulses in both AFC echoes and phase locked echoes function to hold the rephasing process by converting optical coherence into long lived spin coherence.

In conclusion, the atomic coherence conversion process between an optically excited state and a robust spin state used for storage time extension in modified photon echo schemes was analyzed. The analysis demonstrated that the AFC echo belongs to conventional three-pulse photon echoes, while the storage-time control mechanism is based on a rephasing process, such as in phase locked echoes. In quantum memory applications, the AFC technique should be confined only to an extremely weak field limit using single photons as INPUT. Phase locked echo provides efficient retrieval with no limit on single photons, which is good for squeezed light. Although a pencil-like backward scheme of the phase locked echo has a big benefit of quantum imaging and could greatly reduce spontaneous emission noise, applications of single photon-based echoes might be challenging.

Acknowledgments
This work was supported by the CRI program (Center for Photon Information Processing) of the Korean government (MEST) via National Research Foundation.